\documentclass[conference, twocolumn]{IEEEtran}
\usepackage{amsmath,amssymb,graphicx,epsfig,cite,fancyhdr,amsthm,tabulary}
\usepackage{tabulary,bigstrut,array,multirow,booktabs}%
\usepackage{subfigure}
\usepackage{url,hyperref}
\usepackage{tikz}
\usetikzlibrary{positioning,shapes,shadows,arrows,calc,matrix,fit}
\usetikzlibrary{shapes.multipart}
\usepackage{scalefnt}
\usepackage{float}
\usepackage{amsfonts}
\usepackage{multicol}
\usepackage{epstopdf}
\usepackage{footnote}
\usepackage{bm}
\usetikzlibrary{calc}
\usepackage{pdfsync}
\usepackage{multirow}

\makeatletter
\newcommand*\dashline{\rotatebox[origin=c]{90}{$\dabar@\dabar@\dabar@\dabar@\dabar@\dabar@\dabar@$}}
\makeatother

\theoremstyle{remark}

\theoremstyle{definition}

\begin{document}

\title{Improved Modeling of the Correlation Between Continuous-Valued Sources in LDPC-Based DSC}

\author{\IEEEauthorblockN{Mojtaba Vaezi  and Fabrice Labeau}\\
\IEEEauthorblockA{Department of Electrical and Computer Engineering\\
McGill University, Montreal, Quebec H3A 0C6, Canada\\
Email: mojtaba.vaezi@mail.mcgill.ca, fabrice.labeau@mcgill.ca}
}

\maketitle


\begin{abstract}
Accurate modeling of the correlation between the sources plays a crucial role in
the efficiency of distributed source coding (DSC) systems.
This correlation is commonly modeled in the binary
domain by using a single binary symmetric channel (BSC), both for binary and continuous-valued sources.
We show that ``one" BSC cannot accurately capture
the correlation between continuous-valued sources;
a more accurate model requires ``multiple" BSCs, as many as the number of
bits used to represent each sample. We incorporate this new model into
the DSC system that uses low-density parity-check (LDPC) codes for compression.
The standard Slepian-Wolf LDPC decoder
requires a slight modification so that the parameters
of all BSCs are integrated in the log-likelihood ratios (LLRs).
Further, using an interleaver the data belonging to different bit-planes are shuffled
to introduce randomness in the binary domain.
The new system has the same complexity and delay as the standard one.
Simulation results prove the effectiveness of the proposed model and system.
\end{abstract}


\IEEEpeerreviewmaketitle

\section{Introduction}\label{sec:intro}
{\let\thefootnote\relax\footnotetext{This work was supported by Hydro-Qu\'{e}bec,
the Natural Sciences and Engineering Research Council of Canada and McGill
University in the framework of the NSERC/Hydro-Qu\'{e}bec/McGill Industrial
Research Chair in Interactive Information Infrastructure for the Power Grid.}}

Distributed compression of spatially correlated signals, e.g., the observations of neighboring sensors
in high density sensor networks, can drastically reduce
the amount of data to be transmitted.
The efficiency of compression, however, largely depends on the accuracy of the estimation of the correlation between the sources.
The correlation is required at the encoder to determine
the encoding rate; it is also required to initialize the decoding algorithm
in the Slepian-Wolf coding schemes that use channel codes with iterative decoding, e.g., LDPC codes \cite{liveris2002compression}.

The correlation is unknown at the encoder and is modeled by a ``virtual" channel.
The estimation of the {\it virtual correlation channel}
involves modeling it and estimating the model parameter \cite{fang2009correlation,toto2011estimation,cheung2008sampling}.
Therefore, if this virtual correlation channel is not
modeled accurately, even perfect estimation of the model parameter cannot guarantee an efficient compression.

The correlation between the two binary sequences $x^n$ and $y^n$ is
commonly modeled by using a binary symmetric channel (BSC)
with a crossover probability
\begin{align}
p= {\mathrm{Pr}}(y \neq i|x=i), \qquad i\in \{0,1\}.
\label{eq:cor}
\end{align}
The parameter $p$ is either assumed to be known at the encoder \cite{liveris2002compression} 
or needs to be estimated \cite{fang2009correlation,toto2011estimation,cheung2008sampling,varodayan2006rate}.
This model is also widely used in the compression of continuous-valued sources where
Slepian-Wolf coding \cite{SW} is employed to compress the sources after quantization.
Nevertheless, it is known that the correlation between continuous-valued sources
can be modeled more accurately in the continuous domain. Specifically,
the Gaussian distribution and its variations such as the Gaussian Bernoulli-Gaussian (GBG) and the Gaussian-Erasure
(GE) distributions are used for this purpose, particularly when evaluating theoretical bounds
\cite{WZ,bassi2008source,vaezi2011DSC}.

In this paper, we first show that a ``single" BSC cannot accurately model
the correlation between continuous-valued sources, and we propose a new correlation model
that exploits ``multiple" BSCs for this purpose. The number of these
channels is equal to the number of bits used in the binary representation
of one sample. Each channel models the bits with the same significance, i.e., from
the most significant bit (MSB) to the least significant bit (LSB), which is denoted as a bit-plane \cite{shu2012bitplane}.

We next focus on the implementation of the new model
in the LDPC-based compression of continuous-valued sources. We modify the existing decoding
algorithm for this specific model extracted from continuous-valued input sources
and investigate its impact on the coding efficiency. Further, by using an interleaver
before feeding data into the Slepian-Wolf encoder,
the successive bits belonging to one sample are shuffled to
introduce randomness to the errors in the binary domain.
Numerical results, both in the binary and continuous domains,
demonstrate the efficiency of the proposed scheme.

The rest of the paper is organized as follows.
The existing correlation models are discussed in Section \ref{sec:oldcor}.
In Section \ref{sec:newcor} we introduce a new correlation model for continuous-valued sources.
Section \ref{sec:dec} is devoted to integration of the new model to the LDPC-based Slepian-Wolf coding.
Simulation results are presented in Section \ref{sec:sim}.  This is
 followed by conclusions in Section~\ref{sec:sum}.

\section{Existing Correlation Models} 
\label{sec:oldcor}

Lossless compression of correlated sources (Slepian-Wolf coding) is performed through the use of channel codes where
one source is considered as a noisy version of the other one. This requires knowing the correlation between the sources at the decoder.

\subsection{Correlation Between Binary Sources }

The correlation and virtual communication channel between
the binary sequences $x$ and $y$ are the same \cite{stankovic2006code}
and are usually modeled by a BSC with crossover probability $p$.
The parameter of this channel is defined by  \eqref{eq:cor}.
Equivalently, one can obtain $p$ by averaging the Hamming weight of $ x\oplus  y$
for a long run of input data and side information, i.e.,
  \begin{align}
  p=\lim_{n\rightarrow \infty} \frac{1}{n}w_H(x^n \oplus y^n).
  \label{eq:cor2}
  \end{align}
Then, using binary channel coding, near-lossless
compression with a vanishing probability of error can be achieved
provided that the length of the channel code goes to infinity \cite{liveris2002compression,aaron2002compression}.

\subsection{Correlation Between Analog Sources }

In general, the correlation between
the two analog sources $X$ and $Y$ can be defined by
\begin{align}
Y&=X+E, \label{eq:corG1}
\end{align}
where $E$ is a real-valued random variable. Specifically, for
the Gaussian sources we usually have
\begin{align} \label{eq:corG2}
E &  \sim \begin{cases}
{\cal N} (0,\sigma_e^2) \qquad\quad\quad\quad  \text{ w.p. } \quad q_1,\\
{\cal N} (0,\sigma_e^2+\sigma_i^2)\quad\quad  \,\,\,\,\, \text{ w.p. } \quad q_2,\\
 0 \qquad \quad\quad \quad\qquad\quad \;\, \text{ w.p. } \quad 1-q_1-q_2,\\
\end{cases}
\end{align}
in which $\sigma_i^2 \gg \sigma_e^2$ and $q_1+q_2 \leq 1$. This model contains
several well-known models which are suited for video coding and sensor networks.
For example, for $q_1=1$ or $q_2=1$
the Gaussian correlation is obtained, which is broadly used in the literature
when $X$ and $Y$ are Gaussian. Further, for $q_1+q_2=1$
 the GBG and for $q_1+q_2<1$, $q_1q_2=0$ the
 GE models are realized. The latter two models are
 more suitable for video applications \cite{bassi2008source}.
These models are also used for evaluating theoretical bounds and performance limits
\cite{WZ,bassi2008source}.

Although the correlation between continuous-valued sources can be modeled
more accurately in the continuous domain, practically it is usually modeled in the binary domain.
 This is due to the fact that, even for continuous-valued sources, compression is mostly
 done through the use of binary channel codes.\footnote{It is possible to
do compression before quantization; this requires real-number
channel codes and brings about a different paradigm for DSC \cite{vaezi2011DSC}.}
To do so, the two sources are quantized and their correlation is modeled
by a virtual BSC in the binary domain, as shown in Fig.~\ref{fig:model_subfig1}.
In the next section, however, we show that this assumption is not very accurate,
and we propose an alternative, more accurate model.

\section{A New Correlation Channel Model}
\label{sec:newcor}

\subsection{Evaluating the Single BSC Model}
\label{sec:oneBSC}

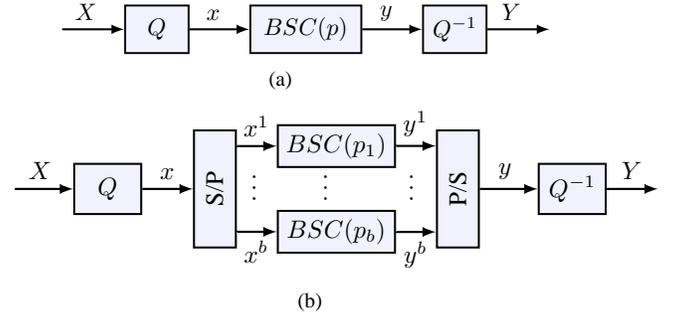
\begin{figure}[t]
\centering
\subfigure[]{
\scalebox {.9}{
\begin{tikzpicture}
[auto, block/.style ={rectangle, draw=blue, thick, fill=blue!5, text width=5em,align=center, rounded corners, minimum height=2em},
 block1/.style ={rectangle, draw=black, thick, fill=blue!5, text width=6.5em,align=center, minimum height=2em},
 line/.style ={draw, thick, -latex',shorten >=2pt},
 cloud/.style ={draw=red, thick, ellipse,fill=red!20,
 minimum height=1em}]
\draw (2.5,-2) node[block1, text width=2em] (C) {$Q$};
\draw (4.7,-2) node[block1,  text width=4em] (C1) {$BSC(p)$};
\draw (6.9,-2) node[block1 , text width=2em] (C2) {$Q^{-1}$};
\draw [-latex] [line width=.3mm] (1.1,-2) |- node {$\qquad X$} (C.west) ;
\draw [-latex] [line width=.3mm] (C.east) |- node {$\qquad x$} (C1.west)  ;
\draw [-latex] [line width=.3mm](C1.east) |- node {$\qquad y$} (C2.west);
\draw [-latex] [line width=.3mm](C2.east) |- node {$\qquad Y$} (8.3,-2) ;
\end{tikzpicture}
}
\label{fig:model_subfig1}
}
\subfigure[]{
\scalebox {.95}{
\begin{tikzpicture}
[auto, block/.style ={rectangle, draw=blue, thick, fill=blue!10, text width=5em,align=center, rounded corners, minimum height=2em},
 block1/.style ={rectangle, draw=black, thick, fill=blue!5, text width=6.5em,align=center, minimum height=1em},
 block2/.style ={rectangle, draw=white, thick, text width=1em,align=center, minimum height=2em},
 line/.style ={draw, thick, -latex',shorten >=2pt},
 cloud/.style ={draw=red, thick, ellipse,fill=red!10,
 minimum height=1em}]
\draw (2.5,-2) node[block1, text width=2em, minimum height=2em] (C) {$Q$};
\draw (4,-2) node[block1 , text width=1em, text height=4em] (C1) {\rotatebox{90}{\;\;\;\, S/P}};
\draw (5.7,-1.4) node[block1, text width=4em] (C5) {$BSC(p_1)$};
\draw (5.7,-2.6) node[block1, text width=4em] (C2) {$BSC(p_b)$};
\draw (7.4,-2) node[block1 , text width=1em, text height=4em] (C3) {\rotatebox{90}{\;\;\;\, P/S}};
\draw (9,-2) node[block1 , text width=2em, , minimum height=2em] (C4) {$Q^{-1}$};
\draw [-latex] [line width=.3mm]  (1.2,-2) |- node {$\qquad X$} (C.west) ;
\draw [-latex] [line width=.3mm] (C.east) |- node {$\qquad x$} (C1.west) ;
\draw [-latex] [line width=.3mm](C1.-63) |- node [below] {$\quad  \;  \; x^b$} (C2.west);
\draw [-latex] [line width=.3mm](C1.63)  |- node {$\quad \; \; x^1$} (C5.west); 
\draw [-latex] [line width=.3mm](C2.east) |- node [below] {$\quad  \;\; y^b$} (C3.243.5);
\draw [-latex] [line width=.3mm](C5.east) |- node  {$\quad  \;\; y^1$} (C3.117);
\draw [-latex] [line width=.3mm](C3.east) |-  node {$\qquad y$} (C4.west);
\draw [-latex] [line width=.3mm](C4.east) |-  node {$\qquad Y$} (10.2,-2);
\draw (4.55,-1.98)node  {\rotatebox{270}{$\hdots$}};
\draw (5.55,-1.98)node  {\rotatebox{270}{$\hdots$}};
\draw (6.75,-1.98)node  {\rotatebox{270}{$\hdots$}};
\end{tikzpicture}
}
\label{fig:model_subfig2}
}
\label{fig:model}
\caption[Optional caption for list of figures]{Virtual correlation channel models for continuous-valued sources ($X$  and $Y$) in the binary domain
  \subref{fig:model_subfig1} Current model.  \subref{fig:model_subfig2} New model for $b$-bit scalar quantizer.
  $x^1$ to $x^b$ are $b$ subsequences of $x$ that contain data belonging to the different bit-planes.
   }
\end{figure}

Let $X$ and $Y$ be two continuous-valued sources.
When using binary channel codes for compression, $X$ and $ Y$
need to be quantized before compression.\footnotemark[\value{footnote}]
Then, as shown in Fig.~\ref{fig:model_subfig1}, the correlation between $x$ and $ y$ (the
binary representation of $X$ and $Y$) is defined in the binary domain
by means of a BSC.

\begin{figure}[!t]
  \centering
 \includegraphics [scale=0.53] {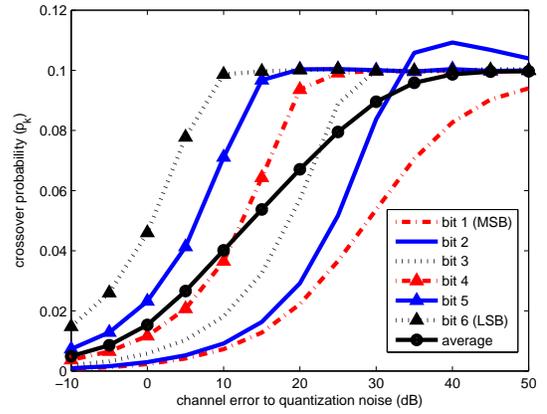}
  \caption{Crossover probabilities of different BSCs,
  each corresponding to one bit-plane, at different channel-error-to-quantization-noise ratio ($\sigma_e^2/\sigma_q^2$).
  $X \sim {\cal N}(0,1)$ and $Y$ is defined by  \eqref{eq:corG1}, \eqref{eq:corG2}
where $q_1=1/5$ and $q_2=0$. Quantization is done using a $6$-bit scalar uniform quantizer.
   }
  \label{fig:BSCs}
  \vspace{-1cm}
\end{figure}

We observe that this model is not very accurate.
This is because the bits resulting from quantization of a sample and
its corresponding side information are not independent. For example,
if $X_i$ (a sample of $ X$) and its counterpart $Y_i$ are the same,
then all bits resulted from those samples will be identical.
That is, the correlation between these bits cannot
be modeled independently.
A more quantitative example is obtained by considering the model in \eqref{eq:corG1} and \eqref{eq:corG2}
with $q_1=1$. Hence, $E  \sim {\cal N} (0,\sigma_e^2)$ and
$\mathrm{Pr}(|E|\geq 2\sigma_e )\leq 5\%$. Now if $\sigma_e =\Delta/2$, where $\Delta$ is
the quantization step size, we will have $\mathrm{Pr}(|E|\geq \Delta )\leq 5\%$. This means that in $y$ (the
binary representation of $Y$), most probably
only the first two lower significant bits will be affected. In other words,
higher significant bits of $x$ and $y$ are similar with high probability.
Numerical results in Fig.~\ref{fig:BSCs} verifies this observation.

The above discussion
indicates that at low channel-error-to-quantization noise ratios ($\sigma_e^2/\sigma_q^2, \sigma_q^2=\Delta^2/12$)
the higher significant bits of $x\oplus y$ (error in the binary domain) are 0, with high probability.
Therefore, correlation parameters differ depending on the bit position (bit-plane); i.e.,
an independent error in the sample (continuous) domain
cannot be translated to an i.i.d. error in the binary domain. Conversely,
a bitwise correlation with a same parameter for all bit positions is not
suited for continuous-valued sources.

In the remaining of this paper, a novel approach is proposed to deal with this problem.
The key is to find a way to effectively model and implement the aforementioned dependency.


\tikzstyle{rect1}=[rectangle,draw=black,fill=white,text centered,text=black,text width=2.8mm,rectangle split,rectangle split parts=#1, draw, anchor=center,minimum height=5.2mm,font=\tiny]
\tikzstyle{rect11}=[rectangle,draw=white,fill=white, text centered,text=black,text width=2.2mm,rectangle split,rectangle split parts=#1, draw, anchor=center,minimum height=5.2mm,font=\scriptsize]
\tikzstyle{branch} = [circle,inner sep=0pt,minimum size=0mm,fill=black,draw=black]
\tikzstyle{line} = [thick]
\tikzstyle{connector} = [draw=black!40,rounded corners]
\tikzstyle{connector2} = [draw=black,rounded corners,rounded corners=3mm]
\tikzstyle{rect2}=[rectangle,minimum size=6mm,text width=.08\textwidth,rounded corners=2mm,text centered,thick,draw=black!60,top color=white,bottom color=blue!15]
\tikzstyle{rect3}=[rectangle,minimum size=6mm,text width=.08\textwidth,rounded corners=2mm,text centered,thick,draw=black!60,top color=blue!10,bottom color=blue!30]
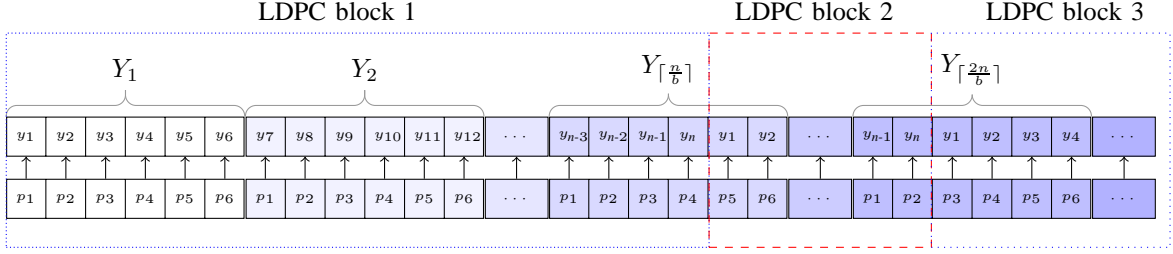
\begin{figure*}[tb]
\begin{center}
\begin{tikzpicture}[scale=.95,auto]
\matrix(C)[row sep=0.3cm,column sep=1cm,xshift=1mm]{
&&& \node (B1) {LDPC block 1};&&&&
\node (B2) {LDPC block 2}; &
\node (B3) {LDPC block 3};\\
};
\matrix(CP)[row sep=0.3cm,column sep=.1mm,yshift=-2.5cm]{
\node(P1) [rect1=6,rectangle split horizontal] { $p_1$\nodepart{two}$p_2$\nodepart{three}$p_3$\nodepart{four}$p_4$\nodepart{five}$p_5$\nodepart{six}$p_6$};&
\node(P2) [rect1=6,rectangle split horizontal,fill=blue!5,] { $p_1$\nodepart{two}$p_2$\nodepart{three}$p_3$\nodepart{four}$p_{4}$\nodepart{five}$p_{5}$\nodepart{six}$p_{6}$};&
\node (Pblank1)[rect1=1,rectangle split horizontal,fill=blue!10,text width=6mm] { $\cdots$};&
\node [name=PN,rect1=6,rectangle split horizontal,fill=blue!15] {$p_1$\nodepart{two}$p_2$\nodepart{three}$p_3$\nodepart{four}$p_{4}$\nodepart{five}$p_{5}$\nodepart{six}$p_{6}$};&
\node (Pblank2)[rect1=1,rectangle split horizontal,fill=blue!20,text width=6mm] { $\cdots$};&
\node(P2N) [rect1=6,rectangle split horizontal,fill=blue!25] {$p_1$\nodepart{two}$p_2$\nodepart{three}$p_3$\nodepart{four}$p_{4}$\nodepart{five}$p_{5}$\nodepart{six}$p_{6}$};&
\node (Pblank3)[rect1=1,rectangle split horizontal,fill=blue!30,text width=6mm] { $\cdots$};\\
};
\matrix(C)[row sep=0.3cm,column sep=.1mm,yshift=-1.2cm]{
\node (LDPC1) {$Y_1$};&
\node (LDPC2) {$Y_2$};&&
\node (LDPCN) {$Y_{ \lceil\frac{n}{b}\rceil}$};&&
\node (LDPC2N) {$Y_{\lceil\frac{2n}{b}\rceil}$};\\
\node(X1) [rect1=6,rectangle split horizontal] {$y_1$\nodepart{two}$y_2$\nodepart{three}$y_3$\nodepart{four}$y_4$\nodepart{five}$y_5$\nodepart{six}$y_6$};&
\node(X2) [rect1=6,rectangle split horizontal,fill=blue!5,] { $y_7$\nodepart{two}$y_8$\nodepart{three}$y_9$\nodepart{four}$y_{10}$\nodepart{five}$y_{11}$\nodepart{six}$y_{12}$};&
\node (Xblank1)[rect1=1,rectangle split horizontal,fill=blue!10,text width=6mm] { $\cdots$};&
\node [name=XN,rect1=6,rectangle split horizontal,fill=blue!15] { $y_{\text {\tiny{\textit{n}-}}3}$\nodepart{two}$y_{\text {\tiny{\textit{n}-}}2}$\nodepart{three}$y_{\text {\tiny{\textit{n}-}}1}$\nodepart{four}$y_{\text {\tiny{\textit{n}}}}$\nodepart{five}$y_1$\nodepart{six}$y_2$};&
\node (Xblank2)[rect1=1,rectangle split horizontal,fill=blue!20,text width=6mm] { $\cdots$};&
\node(X2N) [rect1=6,rectangle split horizontal,fill=blue!25] {$y_{\text {\tiny{\textit{n}-}}1}$ \nodepart{two}$y_{\text {\tiny{\textit{n}}}}$\nodepart{three}$y_1$\nodepart{four}$y_2$\nodepart{five}$y_3$\nodepart{six}$y_4$};&
\node (Xblank3)[rect1=1,rectangle split horizontal,fill=blue!30,text width=6mm] { $\cdots$};\\
};
\draw [connector] (LDPC1) -- +(0,-0.5cm) -| (X1.north west);
\draw [connector] (LDPC1) -- +(0,-0.5cm) -| (X1.north east);
\draw [connector] (LDPC2) -- +(0,-0.5cm) -| (X2.north west);
\draw [connector] (LDPC2) -- +(0,-0.5cm) -| (X2.north east);
\draw [connector] (LDPCN) -- +(0,-0.5cm) -| (XN.north west);
\draw [connector] (LDPCN) -- +(0,-0.5cm) -| (XN.north east);
\draw [connector] (LDPC2N) -- +(0,-0.5cm) -| (X2N.north west);
\draw [connector] (LDPC2N) -- +(0,-0.5cm) -| (X2N.north east);
\draw[blue,densely dotted]  ($(X1.west)+(0,1.45)$) rectangle ($(XN.east)+(-1.1,-1.55)$) ;
\draw[red,dashed]  ($(XN.east)+(-1.1,1.45)$) rectangle ($(X2N.east)+(-2.24,-1.55)$) ;
\draw[blue,dotted]  ($(X2N.east)+(-2.24,1.45)$) rectangle ($(X2N.east)+(1.1,-1.55)$) ;
\draw [connector2, ->] (P1.one north) -- (X1.one south);
\draw [connector2, ->] (P1.two north) -- (X1.two south);
\draw [connector2, ->] (P1.three north) -- (X1.three south);
\draw [connector2, ->] (P1.four north) -- (X1.four south);
\draw [connector2, ->] (P1.five north) -- (X1.five south);
\draw [connector2, ->] (P1.six north) -- (X1.six south);
\draw [connector2, ->] (P2.one north) -- (X2.one south);
\draw [connector2, ->] (P2.two north) -- (X2.two south);
\draw [connector2, ->] (P2.three north) -- (X2.three south);
\draw [connector2, ->] (P2.four north) -- (X2.four south);
\draw [connector2, ->] (P2.five north) -- (X2.five south);
\draw [connector2, ->] (P2.six north) -- (X2.six south);
\draw [connector2, ->] (P2N.one north) -- (X2N.one south);
\draw [connector2, ->] (P2N.two north) -- (X2N.two south);
\draw [connector2, ->] (P2N.three north) -- (X2N.three south);
\draw [connector2, ->] (P2N.four north) -- (X2N.four south);
\draw [connector2, ->] (P2N.five north) -- (X2N.five south);
\draw [connector2, ->] (P2N.six north) -- (X2N.six south);
\draw [connector2, ->] (PN.one north) -- (XN.one south);
\draw [connector2, ->] (PN.two north) -- (XN.two south);
\draw [connector2, ->] (PN.three north) -- (XN.three south);
\draw [connector2, ->] (PN.four north) -- (XN.four south);
\draw [connector2, ->] (PN.five north) -- (XN.five south);
\draw [connector2, ->] (PN.six north) -- (XN.six south);
\draw [connector2, ->] (Pblank1.six north) -- (Xblank1.six south);
\draw [connector2, ->] (Pblank2.six north) -- (Xblank2.six south);
\draw [connector2, ->] (Pblank3.six north) -- (Xblank3.six south);
\end{tikzpicture}
\end{center}
\caption{Variable nodes and their corresponding $p$ in the hybrid LDPC-based decoding for the block length $n=10^4$ and $b=6$.}
\label{fig:LDPC}
\end{figure*}

\subsection{Proposed Model}
\label{sec:parallel}

It is clear that the bits generated from different samples of a source (say $X_i$ and $X_j$) are independent
as long as these samples are generated independently. Also, considering the correlation in
continuous domain, it can be seen that the same argument is valid for the binary representation
of $X$ and $Y$. That is, $x_i$ and $y_j$ are independent if they are generated from different samples.
This is because $X_i$ is related to $Y_i$ (through $E_i$) but it is independent from $Y_j$ for any $j\neq i$.

This indicates that, using a $b$-bit quantizer, $b$ BSCs are enough to efficiently model the correlation between
the two correlated continuous-valued sources; each of these channels
is used to model the correlation between bits corresponding to one bit-plane.
For one thing, $\mbox {BSC}(p_b)$ is used to model the correlation between the MSB's of $X$ and $Y$
in the binary domain.
This is shown in Fig.~\ref{fig:model_subfig2}. Numerical results,
presented in Fig.~\ref{fig:BSCs}, confirm that these channels have different
parameters. Moreover, with high
probability, at low and moderate channel noises we have
\begin{align}
  p_1 \geq  p_2 \geq  \cdots \geq  p_b,
\end{align}
where the indices $1$ to $b$, respectively, represent the channel
corresponding to the LSB to MSB. This is intuitively appealing because
even a small error in continuous domain ($E_i$) can invert the LSB while the MSB
is affected only with large errors.
Note that the parameter of the conventional single BSC model is obtained by
\begin{align}
 p= \frac{1}{b}\sum_{k=1}^{b}p_k.
\end{align}
We next discuss the incorporation of this new model into the
DSC framework that uses LDPC codes for compression.

\section{Decoding Using LDPC Codes}
\label{sec:dec}
In this section, we present three different implementations of the introduced
correlation model in the Slepian-Wolf coding based on LDPC codes.
These are named parallel, sequential, and hybrid decoding.

\subsection{Parallel Decoding}
\label{sec:para}
A first idea is to divide the input sequence into $b$ sub-streams each of which
contains only the bits with the same significance.
Now each channel can be modeled by one BSC with its own parameter.
Hence, we can implement $b$ {\it parallel} LDPC decoders
each corresponding to one correlation channel. This implies
$b$ LDPC decoders at the decoding center, which increases the complexity.
Particularly, effective compression requires codes with different rates, as the
parameter of BSC channel for different bit-planes is different. Then, the code
corresponding to the MSB, for example, will have the highest rate, as it has the smallest $p$.
On the other hand, given a  same code for all channels the MSB will be decoded with
the lowest BER. Given a same LDPC code for all channels,
the complexity increases $b$ times, in the new approach; the delay is the same assuming that
the input of all decoders are available at the reciever.

\subsection{Sequential Decoding}
\label{sec:seq}
By using {\it sequential} decoding, the number of decoders can be reduced to one at the cost of increased delay.
To do so, we let the decoder decode different sub-streams sequentially. Note that each time
the LDPC decoder is initialized with the corresponding $p_k$.
It can be seen that, compared to the parallel decoding, the complexity reduces $b$ times while
the delay increases $b$ times. The latter is due to the fact that in order for decoder to
reconstruct one sample of $X$, it must wait for the output of $b$ LDPC blocks.

\subsection{Hybrid Decoding}
\label{sec:Hyb1}
 A yet more efficient integration of the new correlation model
into the LDPC-based DSC can be achieved just by using a single
LDPC encoder/decoder. This is done in two steps, as explained in the following.

\subsubsection{Manipulating the LLRs}
 The parameters of the multiple-BSC correlation model can be incorporated into
  the LDPC-based DSC by judiciously setting the LLR
sent from (to) the variable nodes. The idea is to take into account
the bit-plane to which each bit belongs.
This requires a slight change in the standard LDPC decoding algorithm. Specifically,
using the notation in \cite{liveris2002compression}, we just need to adjust the LLR
sent from (to) the variable nodes. That is, equation (1) in \cite{liveris2002compression}
will be modified as
\begin{align}
  q_{i,0}=\log \frac{\mathrm{Pr}[x_i=0|y_i]}{\mathrm{Pr}[x_i=1|y_i]}=(1-2y_i)\log \frac{1-p_k[i]}{p_k[i]},
\end{align}
   in which $i=1, \dotsc, n$, $p_k[i] \in \{p_1, \dotsc, p_b \}$, and $k$ represents the bit-plane to which
   $y_i$ (or  $x_i$) belongs. This is illustrated in Fig.~\ref{fig:LDPC}. For example,
   if $x_i$ is the LSB, in its corresponding sample, then $k=1$.
Note that if $b|n$, where $n$ is the code length, then $k=(i \mod b)$.

Since the initial LLR's become more accurate in this method, the number of
iterations required to achieve a same performance reduces. 
However, the performance gap is still noticeable.
To bridge this gap, we propose to interleave the input data (and side information)
in the binary domain.

\subsubsection{Interleaving}
As we discussed in Section~\ref{sec:newcor}, the bits corresponding to each error sample,
which are located in a row, are correlated.
By interleaving $x$ and $y$ before feeding them into the Slepian-Wolf encoder and decoder,
these successive bits can be shuffled 
to introduce randomness to the errors. Then,
it makes better sense to encode data belonging to all bit-planes altogether
as in the conventional approach.
The longer the permutation block input, the more accurate the model and the
better the performance. Interleaving, however, can increase the delay
at the receiver side since we need deinterleaving after
the Slepian-Wolf decoder. To avoid excessive delay, we set the length of interleaving block equal
to the length of the LDPC code.
The improvement in the BER and MSE, only due to interleaving, is remarkably high.
Obviously, we can use interleaving and LLR's manipulation simultaneously; this requires applying interleaving
 to the crossover probabilities, depicted in Fig.~\ref{fig:LDPC}, as well.

Another important advantage of this approach is that it can be used to combat the
bursty correlation channels, as a perfect interleaver transforms a bursty channel
 into an independently distributed channel.
The bursty correlation channel model is capable of addressing the
bursty nature of the correlation between sources in applications
such as sensor networks and video coding, since it takes the memory of
the correlation into account \cite{dupraz2012distributed}.

%
%

\section{Simulation Results and Performance Evaluation}
\label{sec:sim}
\begin{figure*}[t]
\centering
\subfigure[]{
\includegraphics[scale=.53]{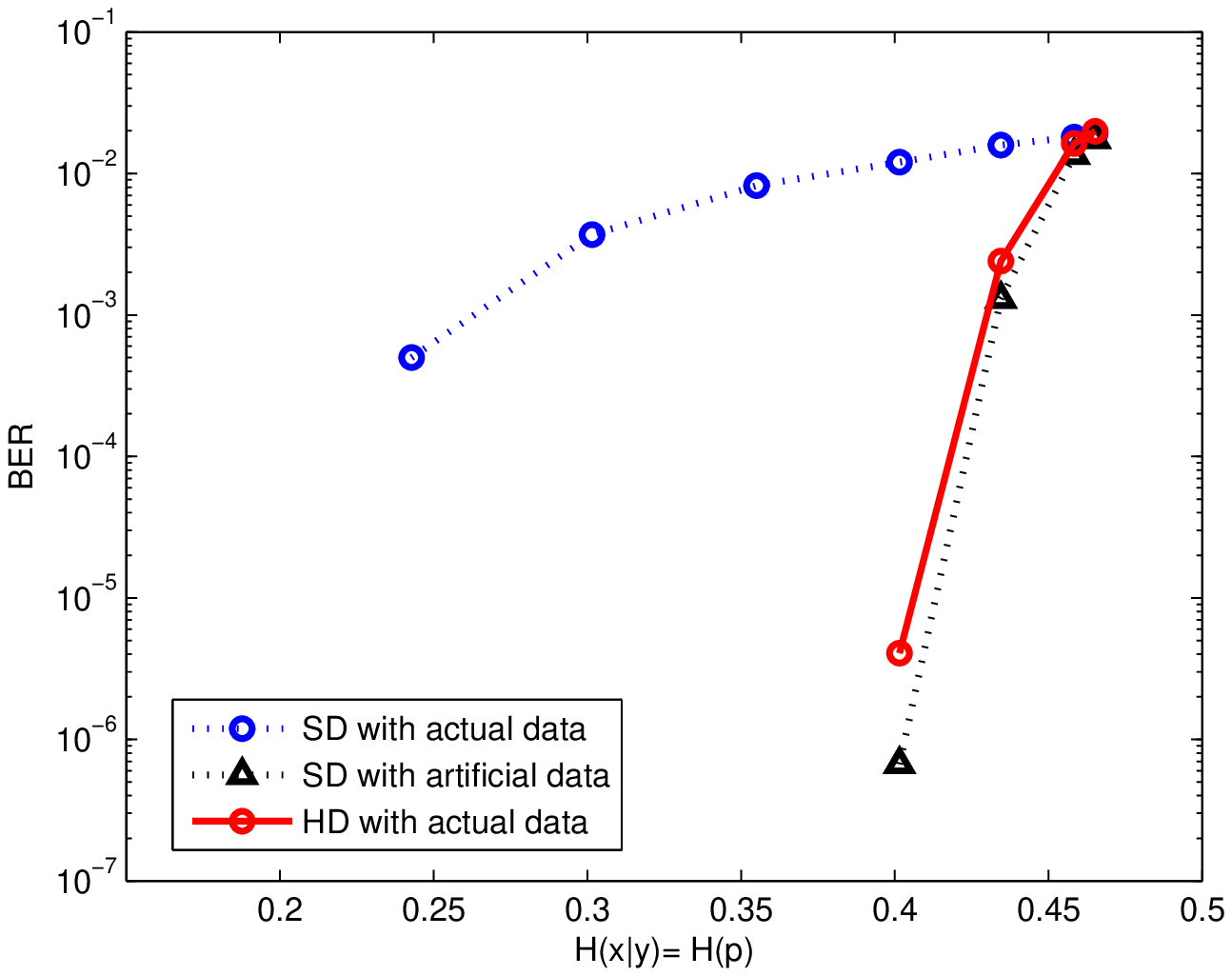}
\label{fig:subfig1}
}
\subfigure[]{
\includegraphics[scale=.53]{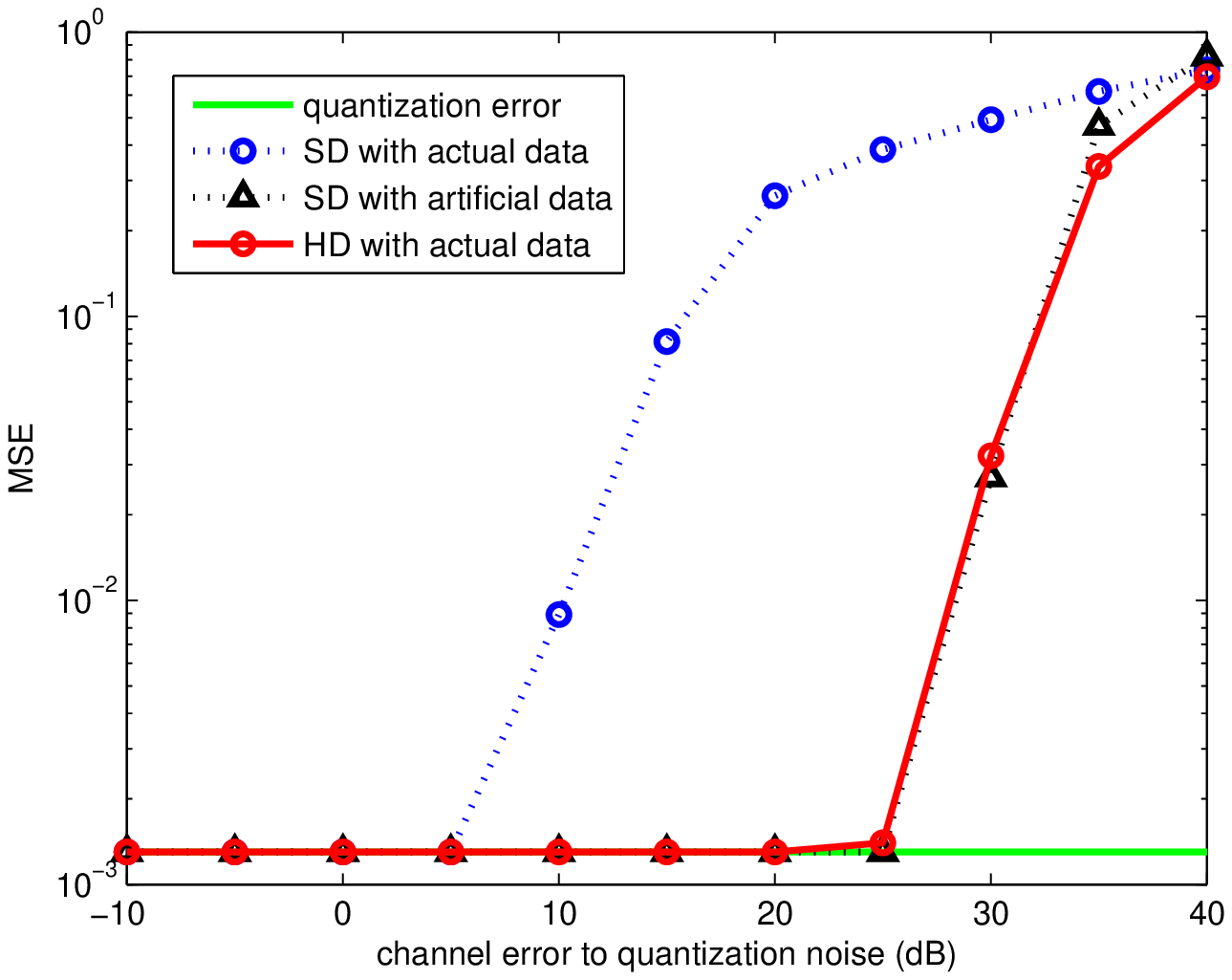}
\label{fig:subfig2}
}
\subfigure[]{
\includegraphics[scale=.53]{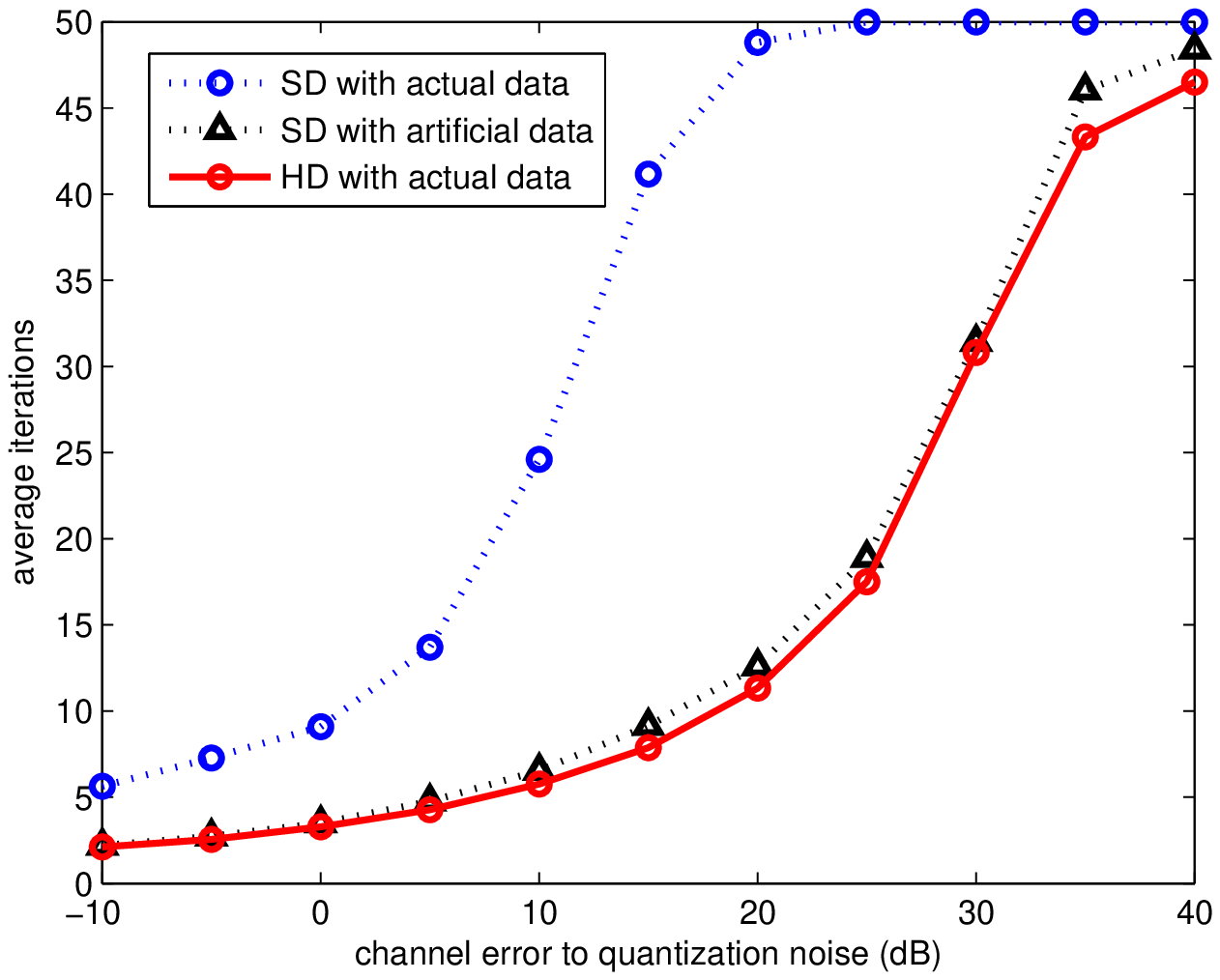}
\label{fig:subfig3}
}
\label{fig:simulations}
\caption[Optional caption for list of figures]{Performance evaluation for irregular rate 1/2 LDPC codes of length $n=10^4$ for maximum iterations of 50.
  SD and HD, respectively, refer to ``standard decoding"
  \cite{liveris2002compression} which is based on single BSC and
  ``hybrid decoding" (proposed in this paper) based on multiple BSCs.
 ``Actual data" is generated by quantizing real-valued $X$ and $Y$ to $x$ and $y$, whereas in
 ``artificial data" $y$ is generated artificially by passing $x$ through a BSC($p$) which is the common approach in the literature.
 \subref{fig:subfig1} The BER performance.  \subref{fig:subfig2} The end to end distortion (MSE).
 \subref{fig:subfig3} Average number of iterations used to achieve the BER and the corresponding MSE in Fig.~\ref{fig:subfig1} and Fig.~\ref{fig:subfig2}.}
\end{figure*}

In this section, we numerically compare the new decoding algorithm
with the conventional approach which considers just one BSC for the correlation model.
We use irregular LDPC code of rate $1/2$ with the degree distribution  \cite{liveris2002compression}
\begin{align*}
\lambda(x)=  0&.234029x + 0.212425x^2 + 0.146898x^5  \\
+&\, 0.102840x^6 + 0.303808x^{19},\\
\rho(x)=  0&.71875x ^7 + 0.28125x^8.
\end{align*}
The frame length is $10^4$ and the bit error rate (BER) and corresponding
mean-squared error (MSE) are measured after 50 itinerations in both schemes.
The source $X$ is a zero mean, unit variance Gaussian. Also the correlation between $X$ and $Y$
is defined by GE channel with $q_1=1/5$, $q_2=0$ in \eqref{eq:corG2}, and channel-error-to-quantization-noise ratio ($\sigma_e^2/\sigma_q^2$)
varies as shown in Fig.~\ref{fig:subfig2}. Both sources are quantized with a 6-bit scalar uniform quantizer.

Simulation results are presented in Fig.~\ref{fig:subfig1}-Fig.~\ref{fig:subfig3}.
In these figures, the ``actual data" refers to the case where binary sequences $x$ and $y$ are obtained from quantizing
$X$ and $Y$. We also compute the BER for
the case that side information $y$ is generated
by passing $x$ through a virtual BSC with parameter $p$, which is
conventional in practical Slepian-Wolf coding \cite{liveris2002compression,fang2009correlation,toto2011estimation,cheung2008sampling,varodayan2006rate}.
This is labeled as  ``artificial data." The fact that
``actual" and ``artificial" side information result in very different BERs,
by itself, indicates that a single BSC is not
an appropriate model for correlation between continuous-valued
sources. On the contrary, the BER resulted from hybrid decoding
with actual side information is significantly better than that of
the conventional approach which shows the suitability of the new model.
  Figure~\ref{fig:subfig2} represents the corresponding MSE.
From these figures, it can be seen that the new scheme (hybrid decoding)
greatly outperforms the existing method, for actual data.
 Furthermore,
as shown is Fig.~\ref{fig:subfig3},
the number of iterations required to achieve such a performance is much smaller than the existing method,
owing to more accurate initial LLRs.

The performance of parallel and sequential decoding, for a same code,
are the same. These schemes benefit from the advantage of working over data
belonging to separate bit-planes. Hence, one BSC can effectively
approximate the corresponding correlation for each bit-plane. Simulation results verify that
separate compression of data belonging to different bit-planes that uses
actual data is as effective as the case that uses artificial side information.
Moreover, there is no need for interleaving. However, an efficient compression,
in parallel and sequential decoding, requires codes with different rates for each bit-plane.
Alternatively, this can be implemented through the use of rate-adaptive LDPC codes
 \cite{artigas2007discover}.
\section{Conclusions}
\label{sec:sum}
We have introduced an improved model for the virtual correlation
between the continuous-valued sources in the binary domain. This model
exploits multiple BSCs rather than the conventional single-BSC model so
that it can deal with the dependency among the bits resulting from quantization of
each error sample by converting the error sequence into multiple i.i.d. sequences.
An efficient implementation of the new model is realized just
by using a single LDPC decoder but judiciously setting the LLR
sent from (to) the variable nodes. The number of
iterations required to achieve the same performance
reduces noticeably as a result of this prudent setting of initial LLRs.
Besides, by interleaving the data
and side information the bits belonging to one error sample are
shuffled which increases the performance of the decoding to a great extent.
   This significant improvement in the BER and MSE is achieved
without any increase in the complexity or delay.
The new scheme can also be used to combat the bursty nature of the
correlation channel in practical applications.


%

\typeout{}

\end{document}